\documentclass[journal]{IEEEtran}

\usepackage[normalem]{ulem}

\usepackage{array}
\usepackage{marvosym}
\usepackage[cmex10]{amsmath}
\usepackage{amssymb,amsthm,epsfig}
\usepackage{graphpap}
\usepackage{color}
\usepackage{epsfig}
\usepackage{amsmath}
\usepackage{latexsym}
\usepackage{amsfonts}
\usepackage{rotating}
\usepackage{multirow}
\usepackage{setspace}
\usepackage{textcomp}
\usepackage{mdwlist}
\usepackage{graphicx}
\usepackage{cite}
\usepackage{supertabular}
\usepackage{longtable}
\usepackage{fullpage,graphicx,url,amsmath,amssymb,mathtools}

\usepackage{subcaption}

\usepackage{soul}
\usepackage{import}

\usepackage{hyperref}
\usepackage{cleveref}
\usepackage{bm}

\usepackage{algorithm}
\usepackage{algorithmicx}
\usepackage{algpseudocode} 
\algnewcommand{\Inputs}[1]{%
  \State \textbf{Inputs:}
  \Statex \hspace*{\algorithmicindent}\parbox[t]{.8\linewidth}{\raggedright #1}
}
\algnewcommand{\Initialize}[1]{%
  \State \textbf{Initialize:}
  \Statex \hspace*{\algorithmicindent}\parbox[t]{.8\linewidth}{\raggedright #1}
}

\usepackage{tikz}
\usepackage{textcomp}

\newcommand{\BEAS}{\begin{eqnarray*}}
\newcommand{\EEAS}{\end{eqnarray*}}
\newcommand{\BEQ}{\begin{equation}}
\newcommand{\EEQ}{\end{equation}}
\newcommand{\BIT}{\begin{itemize}}
\newcommand{\EIT}{\end{itemize}}

\newcommand{\R}{\mathbb{R}}

\newcommand{\comp}{\complement}

\newcommand{\C}{\mathcal{C}}

\newcommand{\B}{\mathcal{B}}

\newcommand{\N}{\mathcal{N}}

\newcommand{\LL}{\mathcal{L}}
\newcommand{\LMP}{\text{LMP}}

\usepackage{tcolorbox}
\usepackage{placeins}

\newcommand{\ones}{\mathbf 1}
\newcommand{\zero}{\mbox{\normalfont\bfseries 0}}
\newcommand{\bg}{\mathbf{g}}
\newcommand{\bMIX}{\mathcal{M}}
\newcommand{\bb}{\mathbf{b}}
\newcommand{\bfl}{\mathbf{f}}
\newcommand{\bd}{\mathbf{d}}

\newcommand{\bs}{\mathbf{s}}
\newcommand{\ba}{\mathbf{a}}

\newcommand{\greg}{g^{(k)}}
\newcommand{\dreg}{d^{(r)}}
\newcommand{\dtot}{d^{(\text{tot})}}
\newcommand{\davg}{\bar{d}^{(\text{tot})}}

\newcommand{\bU}{\mathbf{U}}
\newcommand{\bM}{\mathbf{M}}
\newcommand{\bT}{\mathbf{T}}
\newcommand{\bD}{\mathbf{D}}
\newcommand{\bB}{\mathbf{B}}
\newcommand{\bA}{\mathbf{A}}
\newcommand{\bS}{\mathbf{S}}
\newcommand{\bJ}{\mathbf{J}}
\newcommand{\bE}{\mathbf{E}}
\newcommand{\bI}{\mathbf{I}}
\newcommand{\bP}{\mathbf{P}}
\newcommand{\bX}{\mathbf{X}}

\newcommand{\bL}{\mathbf{L}}
\newcommand{\bG}{\mathbf{G}}
\newcommand{\bLMP}{\mathbf{\text{LMP}}}

\newcommand{\bGamma}{\mathbf{\Gamma}}
\newcommand{\bmu}{\bm{\mu}}
\newcommand{\btau}{\bm{\tau}}
\newcommand{\bpi}{\bm{\pi}}
\newcommand{\bPi}{\bm{\Pi}}
\newcommand{\bY}{\bm{Y}}
\newcommand{\btheta}{\bm{\theta}}

\newtheorem{theorem}{Theorem}[section]

\newtheorem{definition}{Definition}[section]
\newtheorem{assumption}{Assumption}[section]

\usepackage[utf8]{inputenc} 
\usepackage{hyperref}       
\usepackage{url}                 
\usepackage{booktabs}       
\usepackage{amsfonts}       
\usepackage{nicefrac}       
\usepackage{microtype}      
\allowdisplaybreaks
\newcommand\copyrighttext{%
  \footnotesize \textcopyright 2019 IEEE. Personal use of this material is permitted.
  Permission from IEEE must be obtained for all other uses, in any current or future
  media, including reprinting/republishing this material for advertising or promotional
  purposes, creating new collective works, for resale or redistribution to servers or
  lists, or reuse of any copyrighted component of this work in other works.
  }
\newcommand\copyrightnotice{%
\begin{tikzpicture}[remember picture,overlay]
\node[anchor=south,yshift=10pt] at (current page.south) {\fbox{\parbox{\dimexpr\textwidth-\fboxsep-\fboxrule\relax}{\copyrighttext}}};
\end{tikzpicture}%
}

\begin{document}
\title{A Holistic Approach to Forecasting Wholesale Energy Market Prices}
%
%
%

\author{Ana~Radovanovic,
             Tommaso~Nesti, 
             and~Bokan~Chen. 
\thanks{A. Radovanovic and B. Chen are with Google, Inc. Sunnyvale, CA, 94089 (Email: anaradovanovic\MVAt google.com, bokanchen\MVAt google.com).}
\thanks{T. Nesti is with Centrum Wiskunde \& Informatica (CWI), Amsterdam, Netherlands (Email: T.Nesti\MVAt cwi.nl).}
}

\maketitle
\copyrightnotice

\begin{abstract}
Electricity market price predictions enable energy market participants to shape their consumption or supply while meeting their economic and environmental objectives. By utilizing the basic properties of the supply-demand matching process performed by grid operators, known as Optimal Power Flow (OPF), we develop a methodology to recover energy market's structure and predict the resulting nodal prices by using only publicly available data, specifically grid-wide generation type mix, system load, and historical prices.
Our methodology uses the latest advancements in statistical learning to cope with high dimensional and sparse real power grid topologies, as well as scarce, public market data, while exploiting structural characteristics of the underlying OPF mechanism.
Rigorous validations using the Southwest Power Pool (SPP) market data reveal a strong correlation between the grid level mix and corresponding market prices, resulting in accurate day-ahead predictions of real time prices. The proposed approach demonstrates remarkable proximity to the state-of-the-art industry benchmark while assuming a fully decentralized, market-participant perspective. Finally, we recognize limitations of the proposed and other evaluated methodologies in predicting large price spike values. 
\end{abstract}


\IEEEpeerreviewmaketitle


\section*{Nomenclature}
\begin{tabular}{l  p{6.5cm}}
$\N$ & Set of nodes, representing $n$ buses in a power system. \\
$\LL$ & Set of edges, representing $m$ transmission lines.  \\
$\mathcal{G}$ &  $\mathcal{G}=\mathcal{G}(\N,\LL)$ is a connected graph representing the power grid.\\
$\bg$ & Generation vector $\bg \in \mathbb{R}^n$.\\
$\bg^*$ & Optimal generation vector, solution of optimal power flow (OPF) problem.\\
$\bd$ & Demand vector $\bd \in \mathbb{R}^n$. \\
 $C_i(\cdot)$ & The cost function of generation at node $i$. \\
 $\underline{\bg},\bar{\bg}$ & The vectors of generation capacity lower and upper limits, $\underline{\bg},\bar{\bg}\in\mathbb{R}^n$. \\
 \end{tabular}
 \begin{tabular}{l  p{6.5cm}}
 $\underline{\bfl},\bar{\bfl}$ & The vectors of transmission capacity lower and upper limits, $\underline{\bfl},\bar{\bfl}\in\mathbb{R}^m$. \\
 $\bT$ & The power transfer distribution factors matrix (PTDF), $\bT\in\R^{m\times n}$.\\
 $\bA$ & Sub-matrix of the edge-node incidence matrix $\tilde{\bA}$ of $\mathcal{G}$, $\bA\in \R^{m\times (n-1)}$.\\
 $\bB$ & Sub-matrix of the weighted Laplacian matrix $\tilde{\bB}$ of $\mathcal{G}$, $\bB\in\mathbb{R}^{(n-1)\times (n-1)}$.\\
 $\bD$ & Diagonal matrix with $\bD_{\ell\ell}=x_{\ell}^{-1}$, with $x_{\ell}>0$ denoting the reactance of line $\ell\in\LL$, $\bD\in\R^{m\times m}$. \\
 $\lambda$ & The dual variable corresponding to the demand/supply balance constraints in the OPF.\\
 $\tau^-, \tau^+$ & The dual variables corresponding to the generation capacity limit constraints in the OPF.\\
 $\mu^-, \mu^+$ & The dual variables corresponding to the transmission line capacity limit constraints in the OPF.\\ 
 $\bS$ & The congestion matrix (Equation~\ref{eq:congestion_component_S}); $\bs$ denotes one column of the congestion matrix.\\
 $\bMIX$ & The normalized generation and load mix vector we use as inputs to the prediction model (Definition~\ref{def:mix}). \\
 $\dreg(t)$ &  Regional demand in region $r\in\mathcal{R}$ at time $t$, where $\mathcal{R}$ denotes a set of system load regions.\\
 $\greg(t)$ & Total generation from type $k\in\mathcal{K}$ at time $t$, where $\mathcal{K}$ denotes a set of generation types (solar, wind, etc.) in the generation mix data. \\
 $\dtot(t)$ & Total grid wide demand at time $t$. \\
 $\davg$ & Average $\dtot(t)$ across all time intervals $t$.\\
\end{tabular} 
\section{Introduction}
Development of Distributed Energy Resource (DER) technologies enabled the owners of controllable energy assets to shape their wholesale market participation responsively and in a coordinated manner~\cite{Eid2016,Pavic2017}. To address the environmental and operational challenges, besides making the clean energy generation available and cheap, the question remains whether wholesale market prices could be inferred from the supply/demand mix on the grid and, then, used to create a feedback for "shaping" energy asset's production or consumption. In this paper we provide an affirmative response to this inquiry.

Power networks are defined by transmission lines that transport power from generators to loads. Generators and loads are connected to buses, which are commonly referred to as nodes of a power network. Following~\cite{FERC2003}, many wholesale power markets in the US 
adopted the concept of Locational Marginal Prices (LMPs) as electricity prices at the grid nodes. The LMP at a specific node is defined as the marginal cost of supplying the next increment of load at that node, consistent with all power grid operating constraints. 
In this paper, we introduce a novel methodology for predicting LMPs using only publicly available market data.

LMP markets are divided into day-ahead (DA) and real-time (RT, sometimes referred to as intra-day) markets. In the DA market, participants submit bids/offers to buy/sell energy. The Independent System Operator (ISO) then runs the Optimal Power Flow (OPF,~\cite{Huneault1991}) program to derive DA LMPs for each grid's node. The OPF is an optimization problem that determines the generation schedule that minimizes the total system generation cost while satisfying demand/supply balance and network physical constraints~\cite{Schweppe1985}. Since DA scheduled supply may not meet real-time demand, ISOs also calculate RT LMPs every five minutes. 

Our methodology is based on statistical learning techniques that take advantage of the sparsity properties induced by the nature of real grid topologies, underlying physical laws and the resulting OPF solution structure. The emerging field of statistical learning with sparsity~\cite{Hastie2015} aims to utilize sparsity to help recover the underlying signal in a large set of data. Successful applications of sparse machine learning techniques include image/video processing~\cite{Elas2006}, pattern classification~\cite{Huang2007}, face recognition~\cite{Wright2009}, and customers preference 
learning~\cite{Farias2017}. In this paper, using the recent advancements in  compressed sensing~\cite{Donoho2006} and convex optimization~\cite{Tropp2006}, we utilize the OPF problem structure to infer the unknown grid topology, transmission line congestion regimes, and the resulting nodal prices as functions of grid-level generation mix and load. 

We validate the proposed methodology using the Southwest Power Pool (SPP) market data. By focusing on day ahead RT price predictions, we show that the proposed approach has a comparable performance to the state-of-art industry benchmark (Genscape~\cite{Genscape}), which incorporates richer and proprietary information. We further discover that the grid structure and the grid level generation/demand strongly affect the intra-day price shape. However, we also identify the limitations of the newly proposed and other evaluated methodologies to predict large price spike values, even after augmenting the structure-based predictions with statistically estimated error estimates.

The proposed approach assumes a decentralized, market participant-centric perspective, making it fully scalable. To the best of our knowledge, this is the first study that holistically incorporates the structural properties of the grid-level supply-demand matching (OPF), statistical inference and validation using publicly available market data.

In the following two subsections we outline the basics of the power grid modeling, a version of the OPF formulation with its solution structure, as well as the key references in the domain.

\subsection{Power grid and wholesale energy market modeling}\label{ss:gridmodel}

In its full generality, the optimal power flow problem is a nonlinear, nonconvex optimization problem, which is difficult to solve~\cite{Molzahn2013,Bienstock2015strong}. 
For the purpose of this paper, we will focus on a widely used tractable approximation known as DC-OPF~\cite{Sun2010}, which can be formulated as the following optimization problem:
\begin{alignat}{6}
& \underset{\bg}{\min} & & \sum_{i=1}^n C_i(g_i) \label{eq:obj}\\
& \text{s.t.}        & & \ones^T(\bg-\bd)=\zero &   & \qquad :\lambda \label{eq:balance} \\
&					  & & \underline{\bg}\le \bg \le \bar{\bg}\;   &   & \qquad :\btau^-,\btau^+  \label{eq:gen}\\
&					  &	& \underline{\bfl} \le \bT(\bg-\bd)\le \bar{\bfl}\; &\;    & \qquad :\bmu^-,\bmu^+ \label{eq:lines}.
\end{alignat}
The cost functions, $C_i(\cdot)$, $i>0$, are typically modeled using monotonically increasing quadratic or piecewise linear functions~\cite{Glavitsch1991,Sun2010}. Here, we consider generation that has variable and fixed costs of production, but faces no startup, shutdown, no-load costs, or ramping constraints. To that end, we assume $C_i(g_i)=a_ig_i^2+b_ig_i+c_i,\,a_i>0,b_i,c_i\in\R$. Linear and quadratic cost functions, as well as the corresponding generation production range, constitute generators' bids.


The PTDF matrix $\bT$ describes the linear mapping from nodal power injections to active power flows over transmission lines under the  assumption of the Direct Current (DC) approximation~\cite{Purchala2005}. The operators $\le$ and $\ge$ are understood entry-wise.
Following the notation and derivation in~\cite{Kekatos2016}, the PTDF matrix can be written as
\begin{equation}\label{eq:PTDF}
\bT = [\zero\,\, \bD\bA\bB^{-1}],
\end{equation}
where matrices $\bD,\bA,\bB$ describe topological and physical properties of the grid. 
$\bA$ is obtained by deleting the first column of the edge-node incidence matrix $\tilde{\bA}$, which describes which buses are connected to a transmission line; and $\bB$ is obtained by deleting the first row and the first column of the weighted Laplacian matrix $\tilde{\bB}$, which is the matrix representation of the grid graph. The reduced dimension from $n$ to $n-1$ stems from the nullity of the connected grid graph, i.e. $\tilde{\bA}\ones=\zero$. In order to ensure the uniqueness of the optimal solution, without loss of generality, we remove from consideration the node corresponding to the first column, which is selected as the reference bus. In view of the definitions above, matrix $\bA$ is a full-column rank matrix, and $\bB$ is strictly positive definite with non-positive off-diagonal entries. For more detailed derivation and discussion, we refer to~\cite{Kekatos2016}.

Potential generalizations of the above formulation would involve additional operational constraints, such as ramping up/down constraints, power factor constraints, as well as treatments of the reactive power transfer and voltage variation bounds~\cite{Mehta2016}. Nonetheless, in this paper we show that we are able to capture the market structure, as well as the dominant drivers of its dynamics, even under the assumptions of the DC approximation. In addition, the recent advancements in power system technologies and changing regulations (e.g.~\cite{Rule21}) will make the impact of reactive power transfer and the related voltage variations less exaggerated, and the DC-OPF approximation even more accurate.

\begin{figure}[h]
\centering
\includegraphics[width=0.5\textwidth]{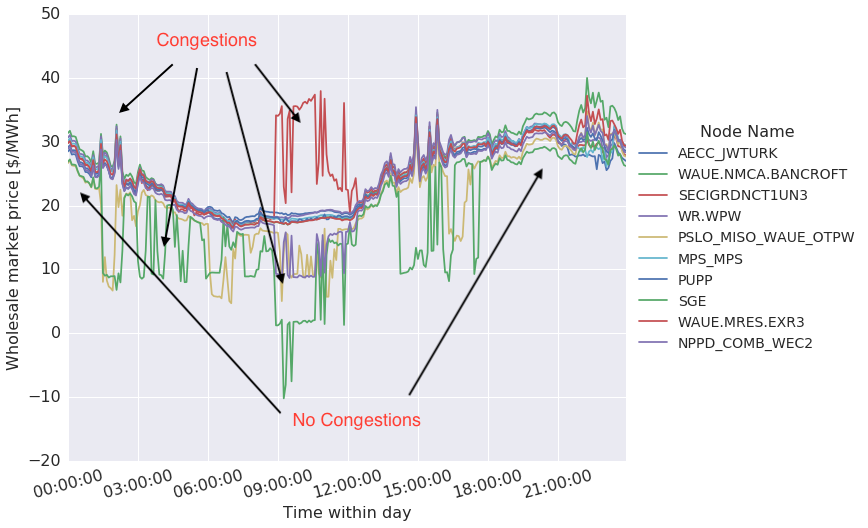}
  \caption{RT market price for randomly selected SPP nodes.}
   \label{SPPRTPrices}
\end{figure}

LMPs are the shadow prices of the real power balance constraints of OPF~\cite{Orfanogianni2007}. More formally, they can be represented as
\begin{equation}\label{eq:LMP_def}
\LMP=\frac{\partial{\mathcal{L}}}{\partial \bd}=\lambda\ones+\bT^\top\bmu,
\end{equation}
where $\frac{\partial{\mathcal{L}}}{\partial \bd}$ denotes the partial derivative of the Lagrangian function of the OPF evaluated at the optimal solution, and $\bmu=\bmu^- -\bmu^+\in\R^m$. The entries of $\bmu$ corresponding to uncongested lines ($\underline{f}_{\ell}<f_{\ell}<\bar{f}_{\ell}$) are equal to zero, while the components corresponding to congested lines are different than zero (in particular, $\mu^+_{\ell}>0$ iff $f_{\ell}=\bar{f}_{\ell}$ and $\mu^-_{\ell}<0$ iff $f_{\ell}=\underline{f}_{\ell}$). As a consequence, if there are no congested lines, all LMPs are equal, i.e. $\LMP_i=\lambda,\;\forall i\in\N$, and the common value $\lambda$ in \eqref{eq:LMP_def} is called the marginal energy component (MEC). The energy component reflects the marginal cost of energy at the reference bus. On the other hand, if some lines are congested, we have $\bmu\ne \zero$ and the LMPs become different (see Figure~\ref{SPPRTPrices}); we call the second term $\tilde{\bpi}=\bT^\top\bmu$ in (\ref{eq:LMP_def}) the marginal congestion component (MCC); in particular, $\tilde{\bpi}_i$ reflects the marginal cost of congestion at bus $i$ relative to the reference bus. 

When ISOs calculate LMPs, they also include the loss component, which is related to the heat dissipated on transmission lines, and is typically negligible compared to the other price components~\cite{SPPreport2016}. For this reason, we omit it from the consideration in this paper, and end up with the marginal energy LMP component (same across all grid nodes) and marginal congestion LMP component, as defined in the previous paragraph and expression (\ref{eq:LMP_def}).

If we recall the definition in (\ref{eq:PTDF}), the marginal congestion price vector (excluding the reference bus) at a given time $t$ can be presented as
\begin{equation}\label{eq:congestion_component}
\bpi(t)=\bB^{-1}\bA^\top\bD\bmu(t)=\bB^{-1}\bs(t)\in\R^{n-1},
\end{equation} 
with $\bs(t)=\bA^T\bD\bmu(t)$. The vector $\bs(t)\in\R^{n-1}$ contains the information on the congested lines, since
\begin{equation}\label{eq:congestion_component_S}
\bs(t)=\sum_{\ell=1}^m \mu_{\ell}(t)x_{\ell}^{-1} \ba_{\ell},
\end{equation} where $\ba_{\ell}\in\mathbb{R}^{n-1}$ is the $\ell$-th column of $\bA^T$. 
The non-zero entries of $\bs(t)$ represent nodes corresponding to congested transmission lines.
Thus, by stacking historical $\bpi(t),\bs(t)$ for $T$ different time intervals as columns of the matrices $\bPi,\bS\in\mathbb{R}^{(n-1)\times T}$, we can rewrite (\ref{eq:congestion_component}) in matrix form as 
\begin{equation}\label{eq:congestion_component_matrix}
\bPi=\bB^{-1}\bS.
\end{equation}
In the following, the matrices $\bB$ and $\bS$ will be referred to as \textit{topology matrix} and \textit{congestion matrix}, respectively.
In Section~\ref{ss:classification}, we use the previous relationship and the properties of matrices $\bB$ and $\bS$ to recover diverse congestion regimes that occur in a grid.

\subsection{Related literature and our contributions}
The previously published work in the same domain can be categorized based on whether it takes an operator-centric or a participant-centric point of view. In the former case, the full knowledge of supply bids, grid topology, and physical properties of the network allow for the explicit computation of nodal LMPs as dual variables of the corresponding OPF optimization. The relevant papers mostly study the impact of uncertainty in total grid load or renewable generation on the resulting prices, while relying on the fact that changes in LMP regimes happen at the so called critical load levels ~\cite{Bo2009,Li2009,Bo2012,Ji2017,Deng2016}. The number of the critical load levels exponentially grows with the size of a grid, making the proposed approaches intractable for use in practice.

The market participant-centric point of view has been much less addressed in the existing literature. A key reference for this work is~\cite{Kekatos2016}, where the authors derive a methodology to recover information on the grid topology based only on publicly available market data, leveraging results from convex optimization and compressed sensing. \cite{Zhou2011} utilizes the structure of the OPF formulation to infer states of transmission lines using only the zonal load levels, without considering generation saturation or grid-level generation mix information. Through the so called System Pattern Regions (SPR), zonal prices are obtained by learning the map between zonal load and the corresponding zonal price, which introduces a large forecasting error. On the other hand, \cite{Geng2016} presents a data-driven approach that exploits structural characteristics by learning nodal prices as a function of nodal loads using support vector machines (SVMs), but is computationally unscalable and limited to synthetically generated, small grid examples.  In~\cite{Birge2017}, the authors propose an inverse optimization approach to estimate the parameters in the OPF, by assuming full knowledge of supply bids, nodal generation and prices, and then obtaining nodal price predictions by solving the OPF with new supply and demand data. The requirement of full knowledge of grid structure makes this approach unusable by market participants.

The contributions of this paper can be summarized as follows:
\begin{description}
\item[1)] By taking market participants' point of view, we propose a price forecasting methodology that utilizes only publicly available data, while enabling users to make more informed bidding decisions.
\item[2)] By combining the theoretical insights on how ISOs derive LMPs, and the latest developments in compressed sensing and machine learning literature, we develop a unique algorithm that can predict market prices with comparable accuracy to the state-of-the-art industry benchmark.
\item[3)] The new approach reveals interesting and potentially very useful insights about different grid state regimes (both in terms of the grid wide generation/load mix, as well as the grid congestions) and their impact on prices across all nodes in the network. It also recovers its limitation to predict some of the price spikes when using only the topology and grid-level generation/demand information, implying that their true cause is of different nature.\end{description}

\section{Energy market structure}
\label{structure}

As Subsection~\ref{ss:gridmodel} suggests, nodal wholesale prices are functions of grid-wide nodal demand and generation. Here, we state the key results from~\cite{Zhou2011}, which formalize the market structure using \textit{pricing regimes} (called \textit{system patterns} in~\cite{Zhou2011}), characterized by grid-wide state vectors that indicate the marginal status of generators and congestion status of transmission lines at optimality. 
Our approach, discussed in Section~\ref{ss:classification}, utilizes these theoretical concepts and parametrizes pricing regimes by a vector of publicly available grid-level generation mix and regional load (called system load) data.


For convenience, we reformulate the OPF problem defined by equations \eqref{eq:obj} - \eqref{eq:lines} as: 
\begin{equation}\label{eq:OPF_MQP} 
\underset{\bg}{\min} \quad  \bJ_1^T \bg +  \frac{1}{2}\bg^T \bJ_2 \bg \qquad  \text{s.t.}         \quad \bA\bg\le \bb+\bE\btheta,   
\end{equation}
where $\bg$ is the optimization variable denoting nodal generation, $\btheta=[\bd\,\,\bar{\bg}]^T$ is a vector of nodal loads and generation capacities, $\bb,\bA,\bE$ are opportunely defined vector and matrices, and $\bJ_1\in\R^n,\bJ_2\in\R^{n\times n}$ define the linear and quadratic costs of generation, respectively.
Note that the objective function \eqref{eq:obj} is a separable quadratic function, 
since the cost of each generator can be assumed independent from other generators' costs~\cite{Glavitsch1991,Sun2010}. Consequently, 
$\bJ_2$ is a positive-definite diagonal matrix.

The formulation in~\eqref{eq:OPF_MQP} allows us to make use of the well-established theory of multiparametric programming~\cite{Tondel2003}, according to which the feasible parameter space of~\eqref{eq:OPF_MQP} can be partitioned into a finite number of convex polytopes, each corresponding to a different pricing regime uniquely defined by the set of marginal generators and congested transmission lines.

More precisely, if $\mathcal{J}$ denotes the index set of constraints of (\ref{eq:OPF_MQP}) and $\bg^*(\btheta)$ the optimal solution for a given parameter vector $\btheta$, we define
\begin{align*}
&\mathcal{B}(\btheta)=\{i\in\mathcal{J}\,\,|\, F_i\bg^*(\btheta)=\bb_i+E_i\btheta\},\\
&\mathcal{B}^\complement(\btheta)=\{i\in\mathcal{J}\,|\, F_i\bg^*(\btheta)<\bb_i+E_i\btheta\}.
\end{align*}
Set $\mathcal{B}$ corresponds to binding (active) constraints, while $\mathcal{B}^\complement$ corresponds to non-binding constraints. Clearly, $\B\cap\B^\comp=\emptyset$ and $\B\cup\B^\comp=\mathcal{J}$. We identify pricing regime 
by the corresponding set of binding constraints. 

The following result, originally established within the multiparametric programming literature in~\cite{Tondel2003}, Theorem 1, and subsequently stated in~\cite{Zhou2011}, Proposition 1, for a variant \footnote{In~\cite{Zhou2011}, the parameter space is constituted by loads $\bd$, while in our approach $\btheta=[\bd,\bar{\bg}]$.} of the LMP forecasting problem, constitutes the theoretical foundations of the methodology presented in this paper.

\begin{theorem}\label{th:main}
Assume that the OPF problem~(\ref{eq:OPF_MQP}) is non-degenerate\footnote{See section 5 in~\cite{Tondel2003} for the definition.}. Then:
\begin{enumerate}
\item The parameter space can be uniquely partitioned into convex polytopes $\Theta(\B_i):=\{\btheta \,:\, \B(\btheta)=\B_i\}$, such that the interior of each polytope corresponds to a unique pricing regime $\B_i$;

\item Within each pricing regime $\B_i$, the optimal generation $\bg^*$ and the associated vector of LMPs~\eqref{eq:LMP_def} are uniquely defined affine functions of $\bd$ and  $\bar{\bg}$.

Overall, the vector of LMPs over the whole parameter space is a continuous, piecewise affine function of nodal demand $\bd$ and generation capacities $\bar{\bg}$.
\end{enumerate}
\end{theorem}

Note that the vector of LMPs is only dependent on $\bar{g_i}$ when generator $i$ becomes saturated, at which point $g^*_i=\bar{g_i}$. Assuming no knowledge on generators' bids, market participants can use $\bg^*$ to learn the piecewise linear mapping of Theorem~\ref{th:main}, as we discuss in Section~\ref{ss:mars}.


\section{Input market data}
\label{data}

The publicly available market data depends on the specific market and commonly includes historical grid-level generation mix, system load, and nodal LMPs. The methodology developed in this paper requires at least the aforementioned components. More granular data (e.g. nodal load and generation) would improve the accuracy of the algorithm, but is not essential.

\begin{figure}[h]      
\centering
    \includegraphics[width=0.5\textwidth]{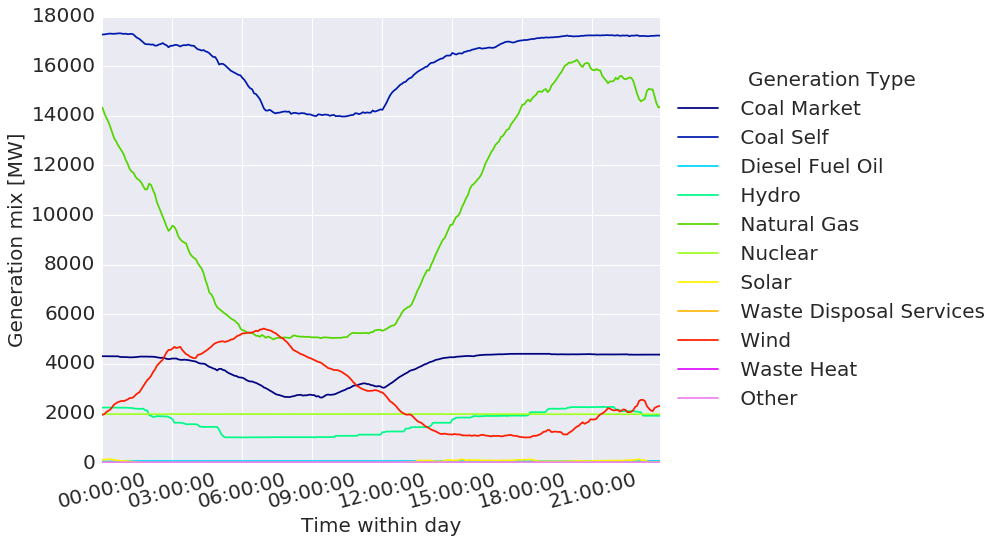}
    \caption{Grid level generation mix.}
    \label{SPPsample}
\end{figure}

In case of the SPP market, historical generation mix is recorded at $5$ minute time granularity and equals to the total average power produced across different types of generation (coal, natural gas, wind, solar, nuclear, etc., see Figure~\ref{SPPsample}). System load consists of regionally aggregated average demand recorded at hourly time granularity.

In addition to the grid level mix and the regional load data, operators release the corresponding nodal RT market prices at $5$ min time granularity. For the purpose of training and validation, in this paper we use six months of SPP data, from June to November of 2017, with $929$ nodes, where we exclude a few new nodal connections with limited price history. 

To facilitate the analysis, we scale the available generation and load data, and refer to the scaled data as $\bMIX$-vectors defined as follows.
\begin{definition}\label{def:mix}
At any time point $t$, $\bMIX(t)$ is a vector that contains three components: normalized generation mix, normalized system load, and scaled total demand.
\begin{itemize}
\item Generation mix is normalized as $\greg(t)/\dtot(t)$,
\item System load is normalized as $\dreg(t)/\dtot(t)$,
\item Total demand is scaled as $\dtot(t)/\davg$,\end{itemize}
where $\dtot(t) = \sum_{r\in\mathcal{R}} \dreg(t) = \sum_{k\in\mathcal{K}} \greg(t)$ is the total demand at time $t$, and
 $\davg$ is the average of $\dtot(t)$ across all time intervals $t$. $\bMIX(t)\in\R^{|\mathcal{R}|+|\mathcal{K}|+1}$ is the vector obtained by concatenating these three components.
\end{definition}
Even though $\bMIX$-vectors are time-indexed, for the reasons of simplicity, in the rest of the paper we omit the time index when referring to $\bMIX(t)$.

\section{New price prediction methodology}
\label{ss:classification}
In this section we propose a novel price prediction methodology that assumes no information on generators' placement, capacities and pricing curves, as well as grid topology, line capacities and load distribution across its nodes, which characterize the DC-OPF in~\eqref{eq:obj}-\eqref{eq:lines}.


The discussion in Section~\ref{structure} provides theoretical concepts for piecewise linear relationships between LMPs and pricing regimes that are uniquely defined by nodal demands and dispatched generation. We utilize this structural property by relating nodal generation and load to the corresponding grid and regional level quantities, 
and introduce the concept of $\bMIX$ regime (Subsection~\ref{ss:mix_regimes}) with the following assumption.

\begin{assumption}
\label{ass1}
Within each $\bMIX$ regime, all generators of the same type (e.g., wind, natural gas, etc.) preserve their production fraction with respect to the total grid level generation of the same type. Similarly, all load within the same geographic region preserves the same consumption ratio when compared to the total load in the region.
\end{assumption}

Intuitively, Assumption~\ref{ass1} states that, within the same intraday $\bMIX$ regime, each generator preserves approximately constant production fraction in relation to the total grid supply of the same type. In other words, when wind generation increases, we assume that all wind generators produce proportionally more power. In case of conventional generators, this translates into generators' typical on/off activity within specific intraday regimes. This simplifying assumption enables us to extend the piecewise affinity in Theorem~\ref{th:main}, and parametrize the pricing regimes using $\bMIX$ vectors. 

More specifically, Theorem~\ref{th:main} establishes that nodal LMPs are piecewise affine function of nodal demand and saturated generation (the parameters in the formulation~\eqref{eq:obj}-\eqref{eq:lines}), which are not publicly available. The introduced proportionality assumption and Theorem~\ref{th:main} result in piecewise affinity of nodal LMPs in grid level saturated generation of each type and regional demands (the $\bMIX$ vectors), within each $\bMIX$ regime. 

In order to learn the LMP function, we conveniently utilize a statistical procedure for fitting continuous adaptive regression splines, called Multivariate Adaptive Regression Splines (MARS)~\cite{Friedman91,MARS}. Relying on the piecewise affinity of the LMP functions, MARS identifies a linear combination of statistically significant truncated spline functions of the form $(x-q)_+$, where $x$ are conveniently scaled covariates (in our case, grid level mix inputs, explained in Subsection~\ref{ss:mars}
), while $q$s are knot locations (price regime switching points) identified by the algorithm.

\subsection{Summary of the forecasting pipeline}
\label{ss:pipeline}

In this section, we summarize the proposed methodology by splitting it into the training and prediction stages.

{\noindent \textbf{Training:}}
\begin{description}
\item[\textbf{Step 0}]: Normalize generation and demand data to generate mix vectors $\bMIX$ (Definition~\ref{def:mix}).

\item[\textbf{Step 1}]: Perform PCA analysis and k-means clustering using the PCA-projected $\bMIX$ vectors to obtain $i = 1,2,\dots,n^\text{Mix}$ $\bMIX$-regimes (Section~\ref{ss:mix_regimes}).

\item[\textbf{Step 2}]: Using historical price data, perform recovery of topology matrix $\bB$ (Section~\ref{ss:topology}).

\item[\textbf{Step 3}]: For each $\bMIX$-regime $i$, compute the corresponding congestion matrix $\bS=\bB^{-1}\bPi$, and run k-means clustering of its columns to obtain the congestion regimes $\C(i)=\{1,2,\ldots,n^\text{Congestion}(i)\}$ (Section~\ref{ss:congestion}).

\item[\textbf{Step 4}]: For each $\bMIX$-regime $i$, relate $\bMIX$-vectors to congestions by training multinomial logistic regression to map $\bMIX$-vectors to congestions $j\in\C(i)$.

\item[\textbf{Step 5}]: For each $(i,j)$, $j\in\C(i)$, and a selected grid node, use MARS to learn LMP deviations as the piecewise linear function of deviations in generation and demand (more details are provided in Section~\ref{ss:mars}).
\end{description}

{\noindent \textbf{Prediction:}}
\begin{description}
\item[\textbf{Step 0}]: Normalize generation and demand forecasts to obtain $\bMIX$-vectors using the same denominators as those used to normalize the training data.

\item[\textbf{Step 1}]: Using the model trained in {\bf Step 1} of the training stage, project the forecasted $\bMIX$-vectors and assign them the matching $\bMIX$-regime $i$.

\item[\textbf{Step 2}]: Within each $\bMIX$-regime $i$, use the trained multinomial logistic regression model in {\bf Step 4} above to assign congestion regime $j\in\C(i)$.

\item[\textbf{Step 3}]: For each $(i,j)$, $j\in\C(i)$, and a selected grid node, map the deviations in generation and load to the resulting price using the model obtained in {\bf Step 5} above.
\item[\textbf{Step 4}]: Perform smoothing of the generated forecasts as described in~\ref{ss:smoothing}.
\end{description}

\begin{figure}[h]    
    \centering  
    \includegraphics[width=0.8\textwidth]{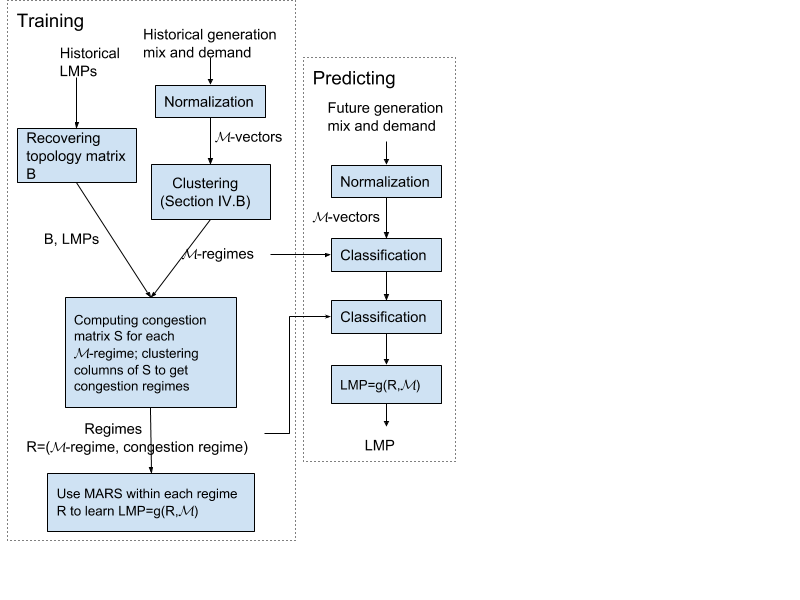}
    \caption{Price prediction pipeline.}
    \label{figure: models}
\end{figure}

The steps and data flows are demonstrated in Figure~\ref{figure: models}. Next, we separately describe each of the modeling components and our approach in validating them.

\subsection{Clustering into $\bMIX$ regimes}
\label{ss:mix_regimes}

\begin{figure}[h]
\centering
    \includegraphics[width=0.4\textwidth]{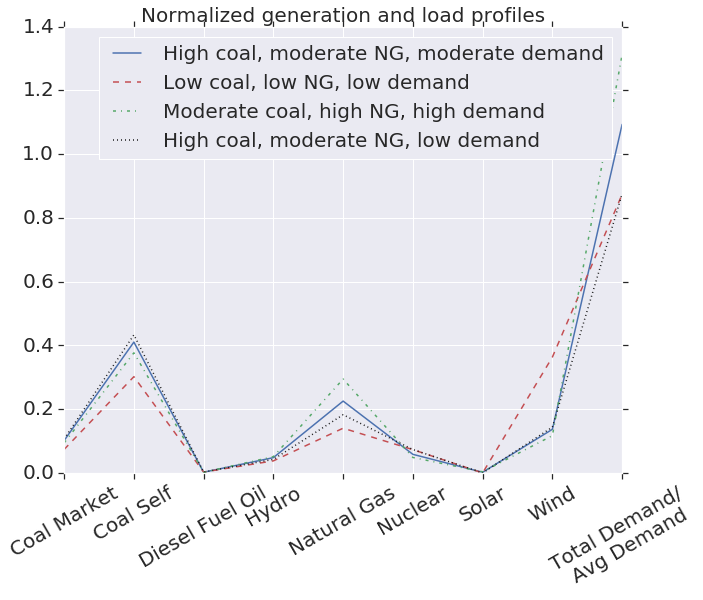}
    \caption{Generation mix corresponding to the centroids of the PCA-based clusters. NG refers to Natural Gas.}\label{fig-pcas}
\end{figure}

\begin{figure}[h]       
\centering
    \includegraphics[width=0.4\textwidth]{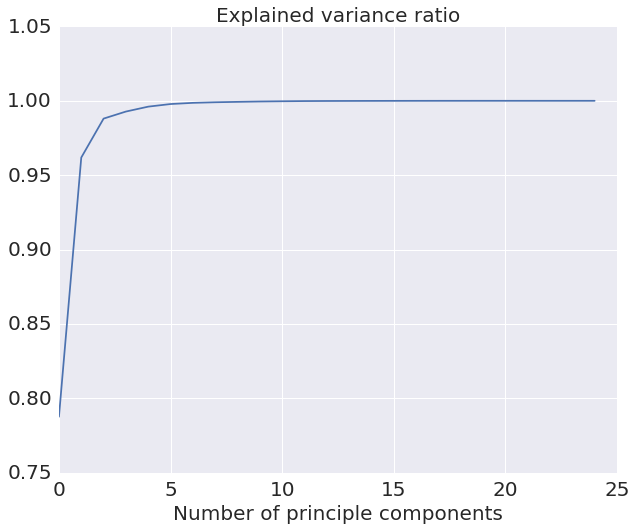}
    \caption{Number of principal components v.s. explained variance ratio.}\label{explained-ratio}
\end{figure}

We classify $\bMIX$-vectors by first applying Principal Component Analysis (PCA,~\cite{HHotelling33}) using $\bMIX$ vectors as defined earlier.
The PCA revealed that only $4$ dominant principal components explain 98\% of the variance (see Figure~\ref{explained-ratio}). Interestingly, the same property is preserved across different time horizons. Then, we perform k-means clustering~\cite{Hartigan79} using the obtained lower dimensional $\bMIX$ representations, where, by applying the fairly standard elbow method, we end up with $4$ $\bMIX$-regimes. Figure~\ref{fig-pcas} shows the centroids of the $4$  clusters, expressed in terms of a subset of the original coordinates corresponding to generation mix. Note that in cases where price exhibits high uncertainty, and when there is enough historical data, one can avoid doing PCA-based clustering, and simply cluster $\bMIX$-vectors based on hour of day. Our validations with the SPP data set show that choosing one versus the other clustering approach does not reveal differences in the final prediction accuracies.


\subsection{Topology recovery}
\label{ss:topology}
The following section is based on the work in~\cite{Kekatos2016}, which focuses of recovering the power grid topology by leveraging only publicly available data. Specifically, the authors derive a methodology to recover matrices $\bB$ and $\bS$ via LMP matrix $\bPi$ and the relation~\eqref{eq:congestion_component_matrix}, based on recent advances in compressed sensing.

For the sake of completeness and reproducibility of our paper, here we
summarize the methodology with the particular focus on 
implementation details, referring the interested reader to~\cite{Kekatos2016} for the complete derivation.

First, note that matrices $\bB$ and $\bS$ enjoy the following structural properties: (i) $\bB$ is a positive definite M-matrix \footnote{See Section 1 in~\cite{Plemmons1977} for the definition of M-matrix.} and is sparse, and (ii) $\bS$ is sparse and low-rank. The sparsity of $\bB$ follows from the fact that the graph underlying a power grid is usually weakly connected (mainly holds for grids inref the USA,~\cite{Babakmehr2015}). The fact that $\bS$ is sparse and low-rank follows from~\eqref{eq:congestion_component_S} and the fact that, almost always, only a very small subset of transmission lines gets congested, implying that most of the terms in the sum in~\eqref{eq:congestion_component_S} are zero.\\
%
In~\cite{Kekatos2016}, the authors suggest to recover matrices $\bB$ and $\bS$ by solving the optimization problem:
\begin{equation}\label{eq:hard_problem}
\begin{aligned}
& \underset{\bB,\bS}{\min}
& & \|\bS\|_{0}+\kappa_0\|\bB\|_{0}\\
& \text{s.t.}
& & \bB\bPi=\bS,\,\bB\succ0,\bB\le\bI,
\end{aligned}
\end{equation}
where $\|\bX\|_{0}$ is the $\ell_0$ pseudo-norm counting the non zero entries of matrix $\bX$, and $\kappa_0\ge 0$.
Since problem~\eqref{eq:hard_problem} is in general NP-hard, the following convex relaxation is used:

\begin{equation}\label{eq:convex_relaxation}
\begin{aligned}
& \underset{\bB,\bS}{\min}
& & \|\bS\|_{1}+\kappa_1\text{tr}(\bP\bB)-\kappa_2\log|\bB| \\
& \text{s.t.}
& & \bB\bPi=\bS,\,\bB\in\mathcal{C},
\end{aligned}
\end{equation}
with $\|\bX\|_1=\sum_{i,j}|\bX_{i,j}|$ denoting the $\ell_1$ norm of matrix $X$, $\bP=\bI-\ones \ones^T,\,\mathcal{C}:=\{\bB:\,\bB\succeq \underline{\zero} ,\, \bB\le \bI\}, \kappa_1,\kappa_2\ge 0$, and $\bB\succeq \underline{\zero}$ denoting a positive semidefinite matrix.

Given that the previous semidefinite program is hard to solve for large grids (order of $10^3$ nodes), we utilize the Alternating Direction Methods of Multipliers (ADMM,~\cite{Boyd10}) to solve the program iteratively. We first replace $\bB$ with three copies $\bB^{(1)},\bB^{(2)},\bB^{(3)}$, yielding the equivalent formulation of \eqref{eq:convex_relaxation}, and then define the matrices $\bM_{12},\bM_{13},\bM$ to be the Lagrange multipliers corresponding to the equality constraints in this new formulation.
%
In every iteration of the ADMM algorithm the variables and Lagrange multipliers are updated by solving appropriate optimization problems, for which closed form solutions are available~\cite{Kekatos2016}. The resulting algorithm is reported in Alg.~\ref{alg:ADMM}.
\begin{algorithm}
    \caption{Topology and congestions recovery.}
    \label{alg:ADMM}
    \begin{algorithmic}[1] 
    \Inputs{
    {\small
    $\bPi,\kappa_1,\kappa_2,\epsilon$
    }
    }
        \Initialize{
        {\small
        $\bB^{(1)}=\bB^{(2)}=\bB^{(3)} = \bI$\\
		$\bS=\bB^{(1)} \bP,\,\bP=\bI-\ones \ones^T$\\
	$\bM^{(12)}=\bM^{(13)}=\underline{\zero}, \bM=\underline{\zero}$}\vspace{0.2cm}
	}
	\While{$\|\bB^{(1)}\bPi-\bS\|_1>\epsilon$}\vspace{0.2cm}
		\State {\small
		$\bB^{(1)} \gets  (\bB^{(2)} - \bM^{(12)} + \bB^{(3)} - \bM^{(13)} +$\\ $\hspace{1.7cm}(\bS-\bM)\bPi^T   - \frac{\kappa_1}{\rho}\bP)(2\bI + \bPi \bPi^{T})^{-1}$
		}
		\State {\small 
		$\bB^{(2)} \gets \min(\bB^{(1)}+\bM^{(12)}, \bI)$}
		\State{\small
		$\bU\bGamma\bU^T=\frac{1}{2}(\bB^{(1)}+\bM^{(13)})(\bB^{(1)}+\bM^{(13)})^T$
		} \label{lst:eigen}
		\State{\small
		$\bB^{(3)} \gets \frac{1}{2}\bU 
		\Bigl(
		\bGamma+(\bGamma^2+4\frac{\kappa_2}{\rho}\bI)^{\frac{1}{2}}
		\Bigr)\bU^T $
		}
		\State{\small
		$\bY_{ij}=\max \Bigl(0,1-\frac{\kappa_2}{\rho |(\bB^{(1)}\bPi+\bM)_{ij}|}\Bigr),$\\ \hspace{1.5cm} $i=1:n-1,j=1:T$
		}\label{lst:max}
		\State{\small
		$\bS  \gets (\bB^{(1)}\bPi + \bM)\odot \bY$
		}\label{lst:hadamard}
		\State{\small
		$\bM^{(12)} \gets \bM^{(12)} + \rho(\bB^{(1)} - \bB^{(2)})$
		}
		\State{\small
		$\bM^{(13)} \gets \bM^{(13)} + \rho(\bB^{(1)} - \bB^{(3)})$
		}
		\State{\small
		$\bM \gets \bM+ \rho(\bB^{(1)}\bPi - \bS)$
		}
	\EndWhile    
	\State {\small $\bB=\bB^{(1)}$}
	\State \textbf{return} $(\bB,\bS)$
    \end{algorithmic}
\end{algorithm}
Matrix $\bU\bGamma\bU^T$ (line~\ref{lst:eigen}) denotes the eigenvalue decomposition of matrix $\frac{1}{2}(\bB^{(1)}+\bM^{(13)})(\bB^{(1)}+\bM^{(13)})^T$, where the maximum operator in (line~\ref{lst:max}) is understood entry-wise and the symbol $\odot$ (line~\ref{lst:hadamard}) denotes the entry-wise product. 

Alg.~\ref{alg:ADMM} is implemented in Python. The parameters $\epsilon$, $\kappa_1,\kappa_2, \rho$ are selected in a way that ensures a reasonable convergence time, while making sure that the resulting matrix $\bB$ has the desired structure (i.e., has all diagonal entries equal to $1$, is sparse, symmetric, with non-positive off-diagonal entries). The selected parameter values are set as $\epsilon=60$, $\kappa_1=1.5$, $\kappa_2=2.0$, $\rho=0.8$. Throughout the execution of the ADMM, we monitor $\ell_1$ norm of matrix $\bB^{(1)}\bPi-\bS$, which monotonically decreases from iteration to iteration. After approximately $1400$ ADMM iterations ($\sim30$ minutes), the monitored $l_1$ norm decreases much slower and, when it reaches $\epsilon=60$, $B$'s entries change insignificantly ($<10^{-4}$) from one iteration to another. Due to the sporadic changes in grid topology, we envision that the topology recovery algorithm runs weekly and, thus, its runtime is not critical for the proposed methodology. 

Since the grid topology is not publicly available, we perform the validation in two ways. Below, we compare the number of nonzero entries obtained by running the topology recovery using the input prices from non-overlapping time intervals (week by week). In order to validate the algorithm's ability to recover the actual links and, consequently, the whole methodology, we run it against the synthetically generated demand and supply inputs, and several IEEE test cases. Our validation process for IEEE 30 bus test case can be found in Appendix. For an extensive testing of the topology recovery accuracy, the interested reader is referred to~\cite{Kekatos2016}.

In practice, nodal connections slowly change due to sporadic repairs and new nodes and, therefore, we expect $\bB$ to be approximately constant. We validate this hypothesis by taking $7$ consecutive weeks of real time market prices and, for each of them, we run the matrix recovery algorithm to infer $\bB_{\text{w},1}, \bB_{\text{w},2}, \bB_{\text{w},3},\dots, \bB_{\text{w},7}$. To evaluate the difference in the recovered links, we first perform entry-wise normalization of all the recovered matrices by dividing each entry with the entry-wise maximum absolute value to obtain scaled matrices $\Hat{\bB}_{\text{w},1}, \Hat{\bB}_{\text{w},2}, \Hat{\bB}_{\text{w},3},\dots, \Hat{\bB}_{\text{w},7}$. Then, we count the identified links by counting off-diagonal entries with absolute values exceeding some given threshold value. 

\begin{figure}[h]    
    \centering  
    \includegraphics[width=0.38\textwidth]{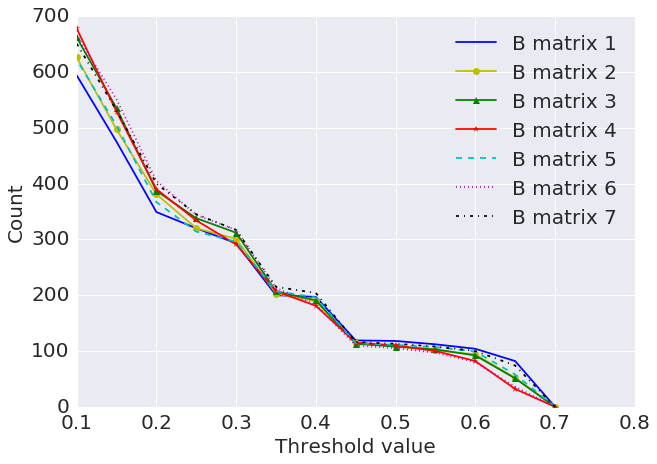}
    \caption{Number of identified transmission lines corresponding to the dominant $\Hat{\bB}$-entry values.}
    \label{links}
\end{figure}

The result of counting the identified grid links for a given threshold across all recovered matrices exhibit a surprising proximity (see Figure~\ref{links}), despite the variable impact of the numerical precision criteria of the topology recovery algorithm, as well as the changing link reactances due to weather conditions and variations in heating induced by the energy transfer. In Table~\ref{table:recovery}, we use $p(\bB_{\text{w},i-1}, \bB_{\text{w},i})$ to express the percent of links identified from week $i$th matrix $\bB_{\text{w},i}$, that are not recovered by week $(i-1)$th matrix $\bB_{\text{w},i-1}$.

\begin{table}
\caption{Differences in recovered links.} \label{table:recovery}
  \centering
  \begin{tabular}{clc}
       & Percent\\
    \hline
  $p(\bB_{\text{w},1}, \bB_{\text{w},2})$  &  6\% \\
  $p(\bB_{\text{w},2}, \bB_{\text{w},3})$  &  6\%\\ 
  $p(\bB_{\text{w},3}, \bB_{\text{w},4})$ & 6\% \\
  $p(\bB_{\text{w},4}, \bB_{\text{w},5})$ & 4\% \\
  $p(\bB_{\text{w},5}, \bB_{\text{w},6})$ & 6\% \\
  $p(\bB_{\text{w},6}, \bB_{\text{w},7})$ & 3\% \\
    \hline
  \end{tabular}
  
\end{table}

\begin{figure}[h]    
    \centering  
    \includegraphics[width=0.4\textwidth]{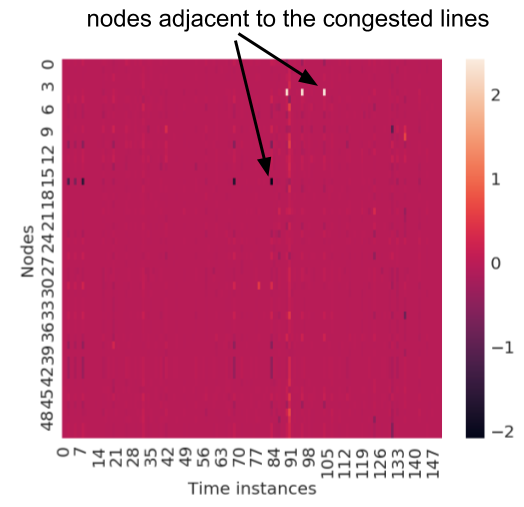}
    \caption{Heatmap of a segment of the congestion matrix.}
    \label{cong-matrix}
\end{figure}

\subsection{Congestion regimes recovery}
\label{ss:congestion}

Apart from the topology matrix $\bB$, the recovery algorithm discussed in Subsection~\ref{ss:topology} allows us to obtain congestion matrix $\bS$, which can be visualized as the heat map in Figure~\ref{cong-matrix}. The x-axis spans time instances, and the y-axis corresponds to node numbers. As can be seen in the figure, most of the entries of the matrix $\bS$ are close to zero, with a few entries having relatively large absolute values. Those entries represent nodes connected to the congested transmission lines.

In contrast to~\cite{Kekatos2016}, in the present work we are mostly interested in recovering \textit{congestion regimes}, defined by the set of congested lines within each $\bMIX$-regime obtained by the previously discussed PCA-based clustering. For each $\bMIX$-regime, we obtain the corresponding matrix $\bS$ using the relationship $\bPi=\bB^{-1}\bS$~\eqref{eq:congestion_component_matrix}, where matrix $\bB$ has previously been obtained as in Section~\ref{ss:topology}.
In view of~\eqref{eq:congestion_component}, we recover congestion regimes by clustering the columns of $\bS$ using again the k-means clustering method. We observe different misclassification fractions depending on the month of the year. During the testing intervals in August and September of $2017$, the misclassification happens in less than $4$\% of instances across all $\bMIX$-regimes, while it can reach $\approx20$\% when the methodology is applied to recover congestions in November 2017. A further investigation of the misclassified instances reveals extremely large price spikes (bursts) in the RT price, which turns out to be the result of the limited information used by the proposed approach.

\subsection{Mapping regime-dependent generation and load to market prices}
\label{ss:mars}

The last stage of the methodology relies on Assumption~\ref{ass1} and Theorem~\ref{th:main}, and learns the piecewise linear function between deviations in generation mix and system load, and deviations in price of the selected node $n$ within the specific regime. 
More specifically, consider regime $(i,j)$, where $i$ is the enumerated $\bMIX$-regime, and $j$ is one of the congestion regimes corresponding to $i$, i.e. $j\in\C(i)$. For each $(i,j)$, we define the corresponding average generation of type $k$ and average regional load $r$ as $\Bar{g}^{(k)}_{(i,j)}$, and $\Bar{d}^{(r)}_{(i,j)}$, respectively. 

Then, without loss of generality, for each regime $(i,j)$, we exploit the piecewise affinity of the LMP price functions and learn local price regime transitions by fitting MARS models~\cite{Friedman91,MARS} to predict nodal price deviations (from the average within the regime) as a function of covariates $\Delta g^{k}_{(i,j)}(t)= g^{(k)}(t) - \Bar{g}^{(k)}_{(i,j)}$, $\Delta d^{r}_{(i,j)}(t)=d^{(r)}(t) - \Bar{d}^{(r)}_{(i,j)}$.

\subsection{Smoothing}\label{ss:smoothing}
Note that we propose to train MARS for each regime $(i,j)$ separately, which can result in observable, small, jitters in the predicted prices at the time instances corresponding to regime shifts. The observed models' miscalibration is more exaggerated when there is a transition to a regime with more frequent price spikes. 

One way to cope with this is to perform smoothing of the raw predicted prices by, first, removing the largest predicted spikes, and, then, by applying interpolation and local smoothing. While the applied smoothing technique recovers the trend in the predicted prices, there is an information loss on the potentially predicted price spikes, which could be critical for risk-aware market participation strategies. 
The performance of the trained models for different training and testing intervals is discussed in the next section.

\section{Performance analysis}
\label{ss:performance}
Even though the proposed approach can be used for predicting any grid node's price, we analyze the performance for two SPP nodes, SPPNORTH and SPPSOUTH hub. The available historical data (beginning of June to end of November 2017), enabled us to test the predictive performance of the new methodology throughout August to November of 2017.

Training and testing is split into four stages triggered by the periodic recovery of the topology matrix $\bB$, which we perform using one week of RT price data from July, August, September, and October.
The corresponding  trained models are tested in the following, non-overlapping, $2$-week chunks at the beginning of August, end of August, end of September, and beginning of November, respectively. The training and predicting within the testing phases follow the steps outlined in Subsection~\ref{ss:pipeline}. 

To train $\bMIX$ and congestion classification models, we typically use $6-8$ weeks of the available historical data, corresponding to the time instances prior to the $2$-week testing phases, while MARS models (Subsection~\ref{ss:mars}) are retrained every day using all available training data for each of the recovered regimes $C(i,j)$. Furthermore, in order to cope with changes in the dispatched generation cost functions, we assign larger weights to more recent training instances.

Note that the real-time execution and evaluation of the proposed methodology is envisioned to be streamlined, where the topology recovery and training of $\bMIX$ and congestion classification models could be repeated biweekly, while MARS models (Subsection~\ref{ss:mars}) are retrained every day prior to providing the forecasts for next day's prices. 

To validate the proposed methodology and investigate its limitations, we evaluate the performance of the following day-ahead forecasts:
\begin{itemize}
\item {\bf ALG-$\bMIX$}: the new methodology applied to the actual generation mix and regional load data within the testing time horizon.
\item {\bf ALG-$\Hat{\bMIX}$}: the new methodology applied to the forecasted generation mix and regional load data (purchased from Tomorrow~\cite{Tomorrow}) within the testing time horizon.
\item{\bf ALG-$\Hat{\bMIX}$ + ARIMA}: forecasts are obtained by adding the estimated hour-of-the-day error term to the ALG-$\Hat{\bMIX}$ forecasts. More specifically, our analysis of algorithm ALG-$\Hat{\bMIX}$'s residuals revealed a strong correlation between the same hours of consecutive days and weak correlation between consecutive hours. We forecast day ahead residuals using ARIMA$(1,1)$ models trained for each hour of day and add them to ALG-$\Hat{\bMIX}$ forecasts.
\item{\bf ALG-$\Hat{\bMIX}$ + Day Ago}: forecasts obtained by adding a day ago ($24$ hours ago) price as an additional feature to the MARS model.
\item {\bf Genscape}: state-of-the-art baseline predictions purchased from Genscape~\cite{Genscape}.
\item {\bf Day Ago}: a naive prediction obtained by using the price from $24$ hours ago as next day's price prediction.
\end{itemize}

In Table~\ref{table:performance}, we included three metrics to evaluate our forecasts: Mean Absolute Percent Error (MAPE), Median Absolute Percent Error (MdAPE), and Root Mean Squared Error (RMSE). Each of the computed metrics has a different sensitivity to the difference between actual and predicted prices. 

\begin{figure}[h]
\centering
\begin{subfigure}[t]{0.39\textwidth}
    \includegraphics[width=\textwidth]{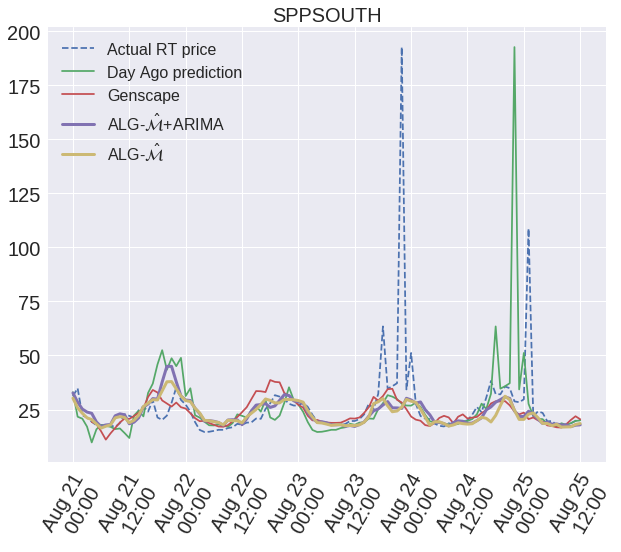}
\end{subfigure}
~
\begin{subfigure}[t]{0.39\textwidth}
    \includegraphics[width=\textwidth]{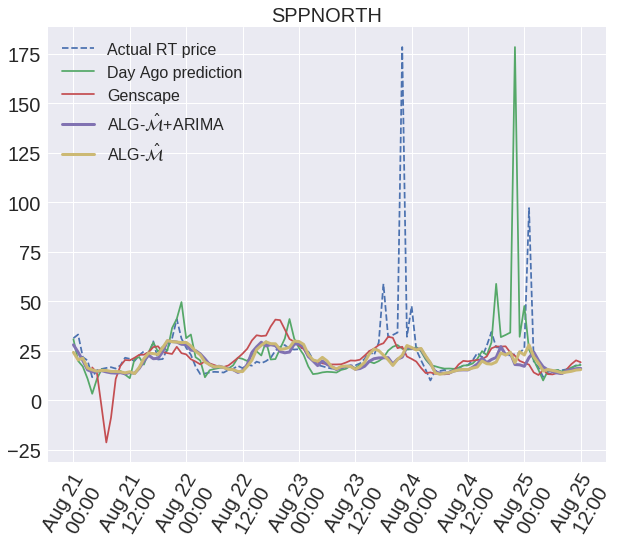}
\end{subfigure}  
\caption{Comparison across available forecasts.}\label{fig:forecast_comparison}
\end{figure}

Based on the evaluation of these metrics, we conclude that the proposed methodology is reasonably accurate throughout the whole testing period, and is robust with respect to the accuracy in day ahead mix forecasts. Figure~\ref{fig:forecast_comparison} shows how well our forecasts follow the trend of actual real time prices. 
Furthermore, we observe that the proposed approach has a comparable performance
to the industry benchmark, Genscape, with respect to all of the considered metrics, which is remarkable given that Genscape incorporates richer and proprietary data. 

\begin{figure}
\centering
\includegraphics[width=0.39\textwidth]{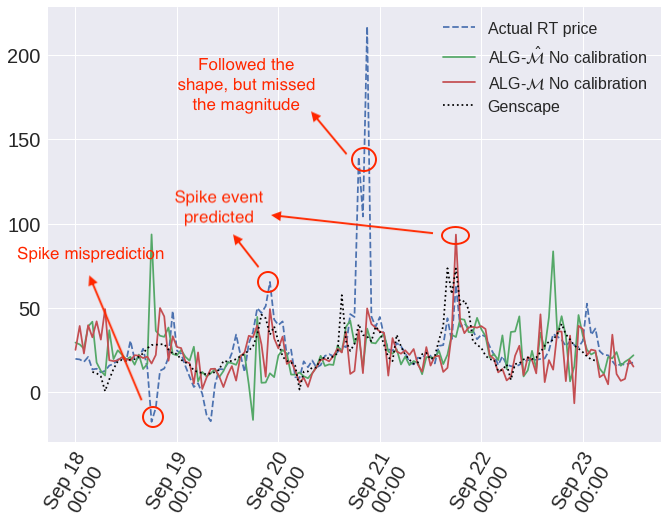}
\caption{Spike prediction.}\label{fig:calibration}
\end{figure}

While the smoothing discussed in Subsection~\ref{ss:smoothing} helps to more accurately predict the trend in day ahead RT prices, we lose valuable spike information. As Figure~\ref{fig:calibration} shows, the original, uncalibrated, ALG-$\Hat{\bMIX}$ forecasts are capable of predicting spike events caused by the grid level generation and demand mix.
The spike predictions could be utilized by market participant to build risk-averse trading strategies, potentially avoiding large losses.
However, predicting price spike \textit{magnitudes} turns out to be a more challenging task, and some of this difficulty can certainly be attributed to the accuracy of the grid level mix predictions. 

Finally, we note that there are spikes that the newly proposed methodology does not predict, even when the absolute knowledge of the grid level mix is available.
The failure to predict such spikes suggests that they are the result of some other, unknown phenomena, not captured by the available data. Possible causes include short-term infrastructure failures, as well as limitations related to using a simple version of the DC-OPF formulation, without taking into account reserves and ramping constraints.

Since the process of supply-demand matching for DA energy market involves solving the same optimization problem, the models trained for the RT price prediction can be used to infer DA prices as well, while the inputs in this case would correspond to the day-ahead cleared generation and system load, and is typically less variable (i.e., we can expect smaller prediction errors).

\begin{table}[h]
\caption{Forecast Accuracy Comparison For All Prices.} \label{table:performance}
\centering
\begin{tabular}{c| c| c| c |c}
Node                                                                &  Approach                                  & MAPE  & MdAPE       & RMSE \\ \hline\hline
\multirow{5}{*}{\rotatebox{90}{SPPSOUTH}}   & ALG-$\bMIX$                             & $26.4\%$    &  17.4\%                        &  19.6        \\ 
                                                                         & ALG-$\hat{\bMIX}$ 			& $25.4\%$     &  15.8\%                       &  19.4       \\
                                                                         & ALG-$\hat{\bMIX}$+ARIMA       & $26.3\%$    &   15.5\%                       &  19.9       \\
                                                                         &ALG-$\hat{\bMIX}$+Day Ago      & $26.0\%$    &  14.9\%                        &  19.3       \\
                                                                         & Genscape                                  & $21.7\%$    &   11.2\%                       &  19.2       \\
                                                                         & Day Ago                                     & $31.1\%$    &   12.9\%                      &   23.5      \\ 
                                                                         \hline
\multirow{5}{*}{\rotatebox{90}{SPPNORTH}}   & ALG-$\bMIX$                             & $39.2\%$    &    17.1\%                      &  18.5        \\ 
                                                                         & ALG-$\hat{\bMIX}$ 			 &$36.9\%$    &     17.4\%                     &   18.0       \\
                                                                         & ALG-$\hat{\bMIX}$+ARIMA       & $38.0\%$    &     17.7\%                      &   18.4       \\
                                                                         &ALG-$\hat{\bMIX}$+Day Ago      & $37.1\%$   &       16.6\%                   &    18.3     \\
                                                                         & Genscape                                  & $28.2\%$    &     13.7\%                     &    19.1       \\
                                                                         & Day Ago                                     & $50.0\%$    &      16.2\%                    &     23.4       \\ \hline                                                                                                                             
\end{tabular}
\end{table}

\section{Conclusions and future work}
In this paper we show that the wholesale energy market structure can be inferred using limited, publicly available, historical market data (grid level generation type mix, system load mix and nodal prices). By utilizing the basic underlying physical model that captures generation-load matching on the grid, we develop a methodology for predicting nodal market prices. 
Extensive and rigorous validations using the Southwest Power Pool (SPP) market data
shows that the price values are significantly impacted by this basic structural information. 

The proposed approach, in its basic (ALG-$\Hat{\bMIX}$) and augmented forms (ALG-$\Hat{\bMIX}$ + ARIMA, ALG-$\Hat{\bMIX}$ + Day Ago), fairly closely matches the performance of the state-of-the-art baseline (Genscape~\cite{Genscape}) in predicting day ahead RT price trend, using only grid level, publicly available data. 

Furthermore, a derivative of the proposed methodology is the uncalibrated ALG-$\Hat{\bMIX}$ forecast, which identifies spike events related to generation and demand mix changes and turns out to be a unique feature across the considered algorithms. This capability enables market participants to design trading strategies that will protect them from potentially large market losses.

The proposed methodology can be enhanced with additional data. For example, the statistical models in this paper did not capture seasonality patterns in generation and load, as well as the impact of other relevant data, such as available reserves and their prices, or ramping constraints. To that end, we believe that using the additional data sources can only improve the prediction performance, and it will be part of our future investigations. The other potential area of exploration involves studying the impact of localized measurements at market participants' sites to improve the corresponding local market predictions. After all, it is expected that each market participant's goal is to maximize its own financial reward while reducing environmental impact, and improving market predictions is the key instrument for achieving these objectives.


\section*{Acknowledgment}
The authors would like to thank Thomas Olavson and John Platt for their support and constructive comments, as well as the editor and reviewers for their many useful suggestions.

\bibliographystyle{IEEEtran}
\bibliography{IEEEabrv,LMPPredictions-NIPS2018}

\section*{Appendix: IEEE 30-Bus System Case Study}
In this section we report the results for IEEE 30-bus test system, which is taken from the MATPOWER toolbox (case 30,~\cite{Matpower2011}). Due to the small number of generators, we do not normalize the generation/load values to obtain mix vectors, and there is no need to conduct the PCA analysis. The test case includes all parameters needed to run DC-OPF, but since it does not include renewable generators, we modify a subset of the generators (specifically generators $2$ and $4$) into renewable generators as follows: we set the corresponding cost functions to be equal to $0$, and we allow the generation limit $\bar{g}$ to vary according to a distribution learned from CAISO data~\cite{CAISO2}, as explained below. To make sure that we get a few congested lines, we also uniformly multiply the transmission line limits by a small factor, set at $1.2$.

For demand profiles, we download $3$ months of historical total demand profiles from S$\&$P Global~\cite{SP}, collected at \textit{hourly granularity}. We then fit the multivariate normal distribution to the historical data and obtain model $\dtot\sim N_{24}(\mu_\text{demand}, \Sigma_\text{demand})$ from which we can sample daily total load profiles. We scale down the total demand so that its mean value matches the base-level demand of the test case. Finally, for each load node $i$, we assign its demand to be the fixed fraction of the total, grid-level, demand as $\alpha_i$, where $\sum_i \alpha_i = 1, \alpha_i >0$. 

To simulate renewable generation profiles, we download a typical daily profile for PV generation in the CAISO market from California ISO, which we denote as $\tilde{G}\in\mathbb{R}^{24}$. A daily realization $G\in\R^{24}$ is then obtained by sampling from normal distribution $G_h \sim N(\tilde{G_h}, \tilde{G_h}\sigma^2_\text{supply})$ for each hour $h$, where we set $\sigma^2_\text{supply}=0.1$. The generated profiles are scaled down to be consistent with the other generators. Finally, similarly as in the case of load nodes, renewable generation for the specific nodes are obtained by fixing the ratios to the simulated, grid-level, generation, where these fractions are expressed by $\beta_k$, where $\sum_k\beta_k = 1, \beta_k >0$. 

Finally, we run the DC-OPF in MATPOWER to obtain LMPs, which we use for training and testing of our algorithm. The training and testing periods consist of $3$ months and $50$ days, respectively.

\begin{figure}[h]
\centering
\begin{subfigure}[t]{0.226\textwidth}
    \includegraphics[width=\textwidth]{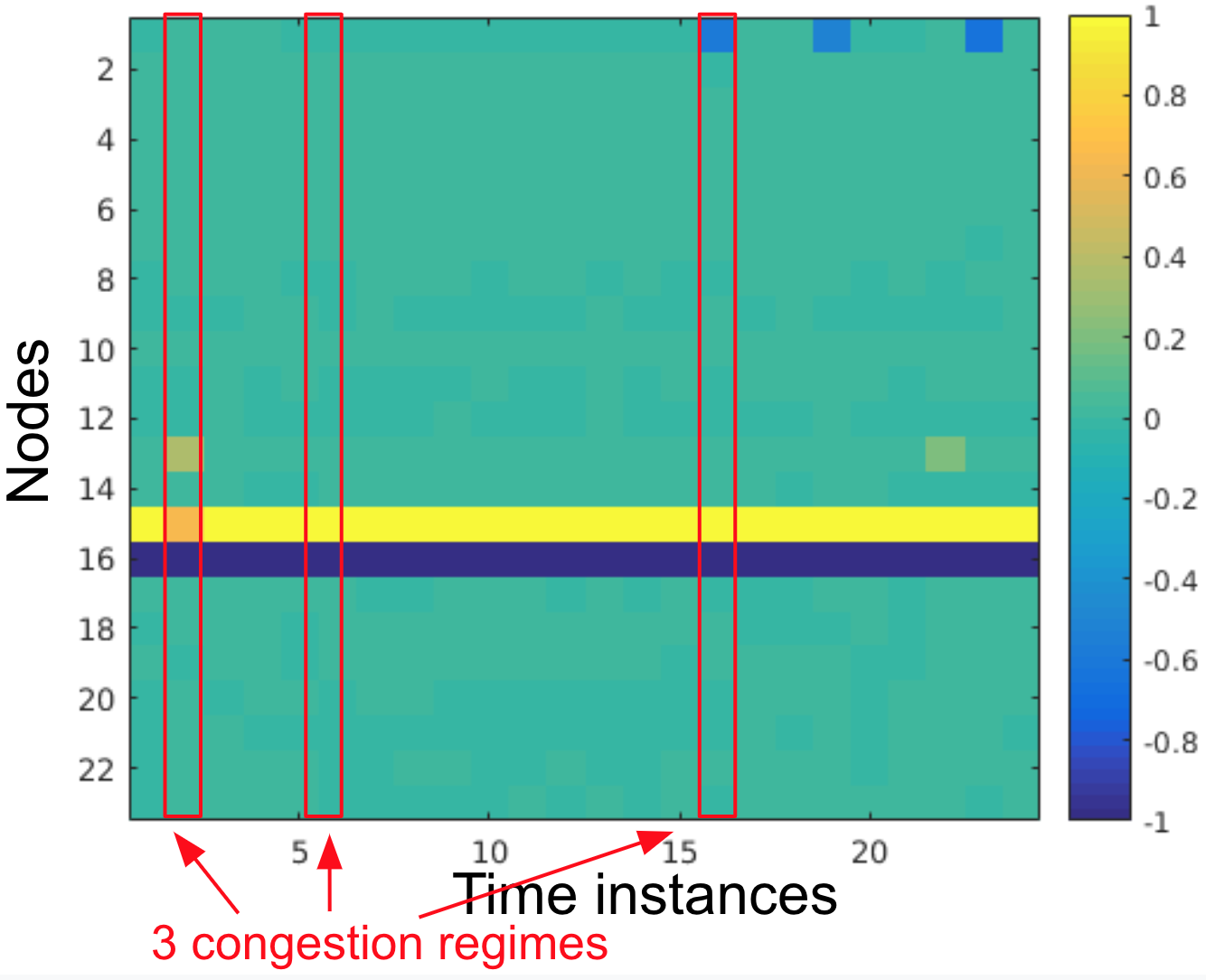}
    \caption{Actual}
\end{subfigure}
~
\begin{subfigure}[t]{0.24\textwidth}
    \includegraphics[width=\textwidth]{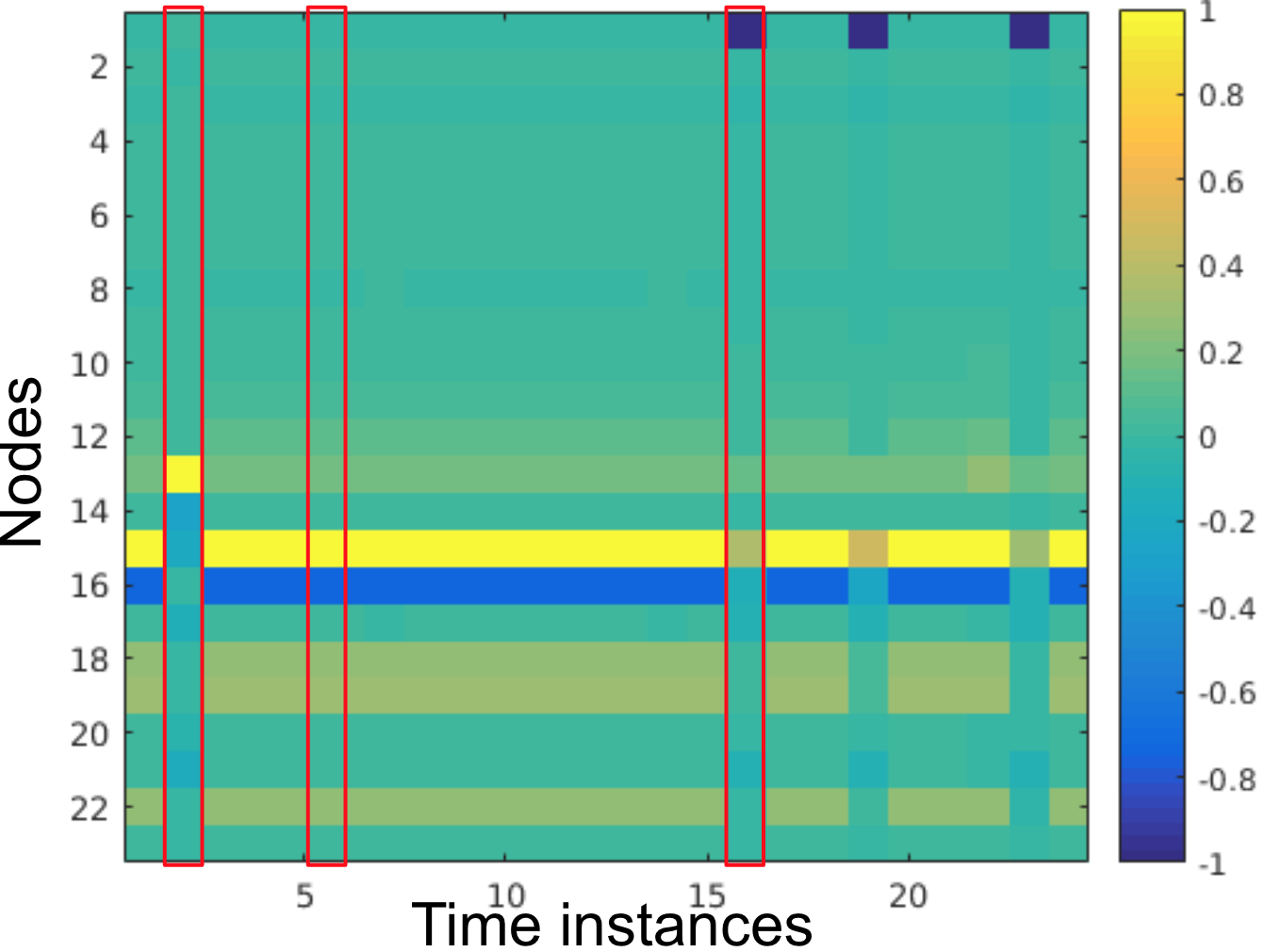}
    \caption{Recovered}
\end{subfigure}  
\caption{Comparison between Actual and Recovered Congestion.}{\label{fig:comp}}
\end{figure}

The topology recovery method described in this paper can only recover transmission lines that get affected by congestions. The focus of our validations was to check the accuracy when recovering the congestion matrices. In Figure~\ref{fig:comp}, we show the comparison between snapshots of the actual and recovered congestion matrices. The red boxes highlights three of the dominant congestion regimes. The graph shows that the algorithm is able to recover the congestion matrix with a reasonable accuracy and, as a result, identify the congestion regimes.

To evaluate the predictive performance of our approach, we use a synthetically generated day-ahead forecasts for total load and total renewable generation. First, we generate $100$ samples from distributions $\dtot(i)\in\R^{24}$ and $G(i)\in\R^{24}$, for day $i$, and over a testing period of $50$ days. Then, we cluster these samples into typical profiles ($5$ for demand, and $2$ for renewable supply), and match each sample path to the closest identified typical profile (in terms of L2-norm) to obtain the synthetic day-ahead forecasts $\widehat{\dtot}(i)$ and $\widehat{G}(i)$ for each day. 

By stacking $\dtot,\widehat{\dtot}(i),G,\widehat{G}$ for the different days in the testing period as columns of the matrices $\bL,\hat{\bL},\bG,\hat{\bG}\in\R^{24,50}$, 
the forecasting errors are expressed by
\[
\text{err}^{(\text{demand})}=\frac{\|\bL-\hat{\bL}\|_F}{\|\bL\|_F},\,
\text{err}^{(\text{ren.gen.})}=\frac{\|\bG-\hat{\bG}\|_F}{\|\bG\|_F},
\]
where $\|A\|_F=\sqrt{\sum_{i,j}|a_{ij}|^2}$ denotes the Frobenius norm of matrix $A$. Similarly, for each node $k$, we stack the actual and predicted LMPs as columns of the matrices $\bLMP_k,\widehat{\bLMP}_k\in\R^{24,50}$. 

Then, the predictive performance of our methodology is evaluated using the mean relative error $\text{err}_k$ across all nodes $\frac{1}{n}\sum_{k=1}^n \text{err}_k$,
where
\[
\text{err}_k=\frac{ \|\bLMP_k-\widehat{\bLMP}_k\|_F }{\|\bLMP_k\|_F}.
\]

\begin{figure}[h]
\centering
\begin{subfigure}[t]{0.35\textwidth}
    \includegraphics[width=\textwidth]{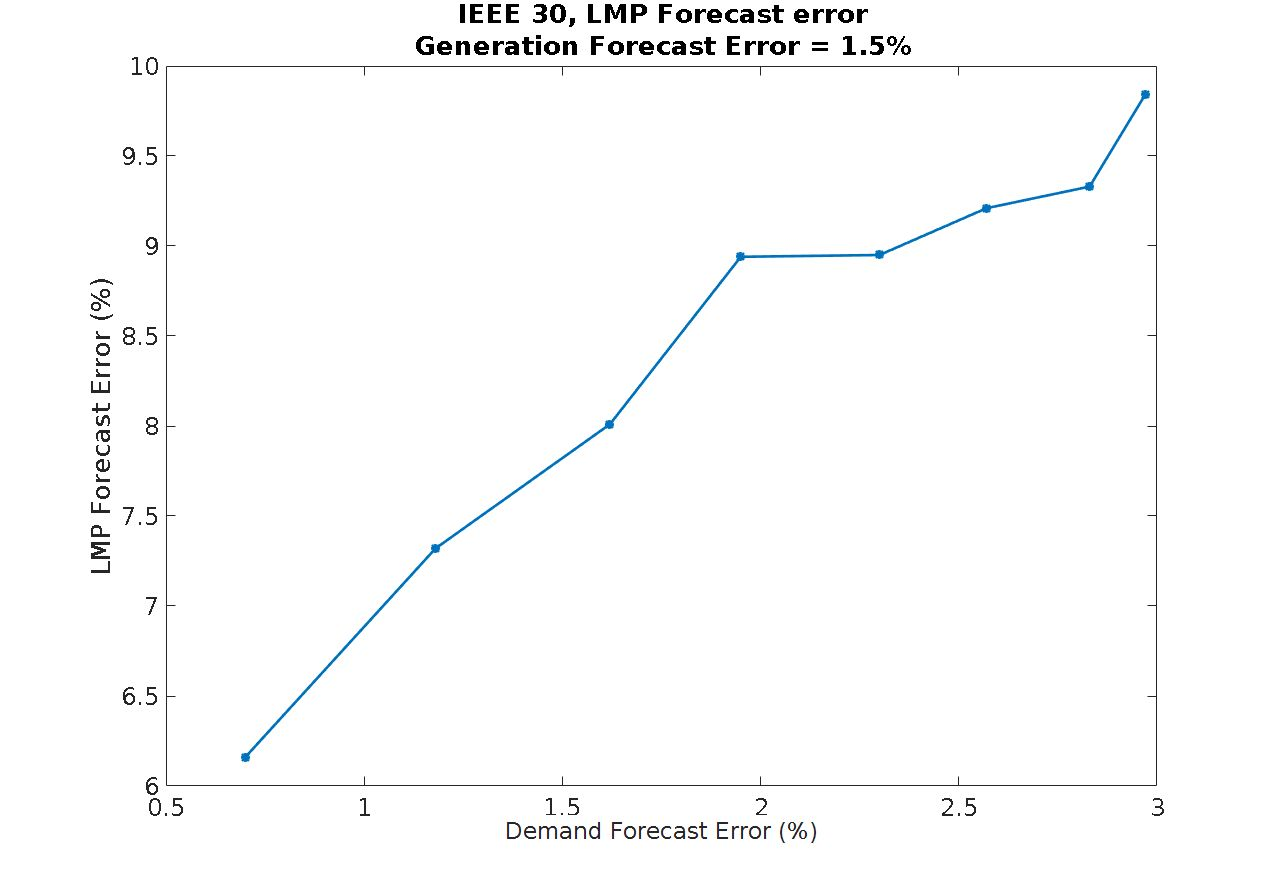}
    \caption{Renewable generation forecast error fixed at $1.5\%$.}
\end{subfigure}
~
\begin{subfigure}[t]{0.35\textwidth}
    \includegraphics[width=\textwidth]{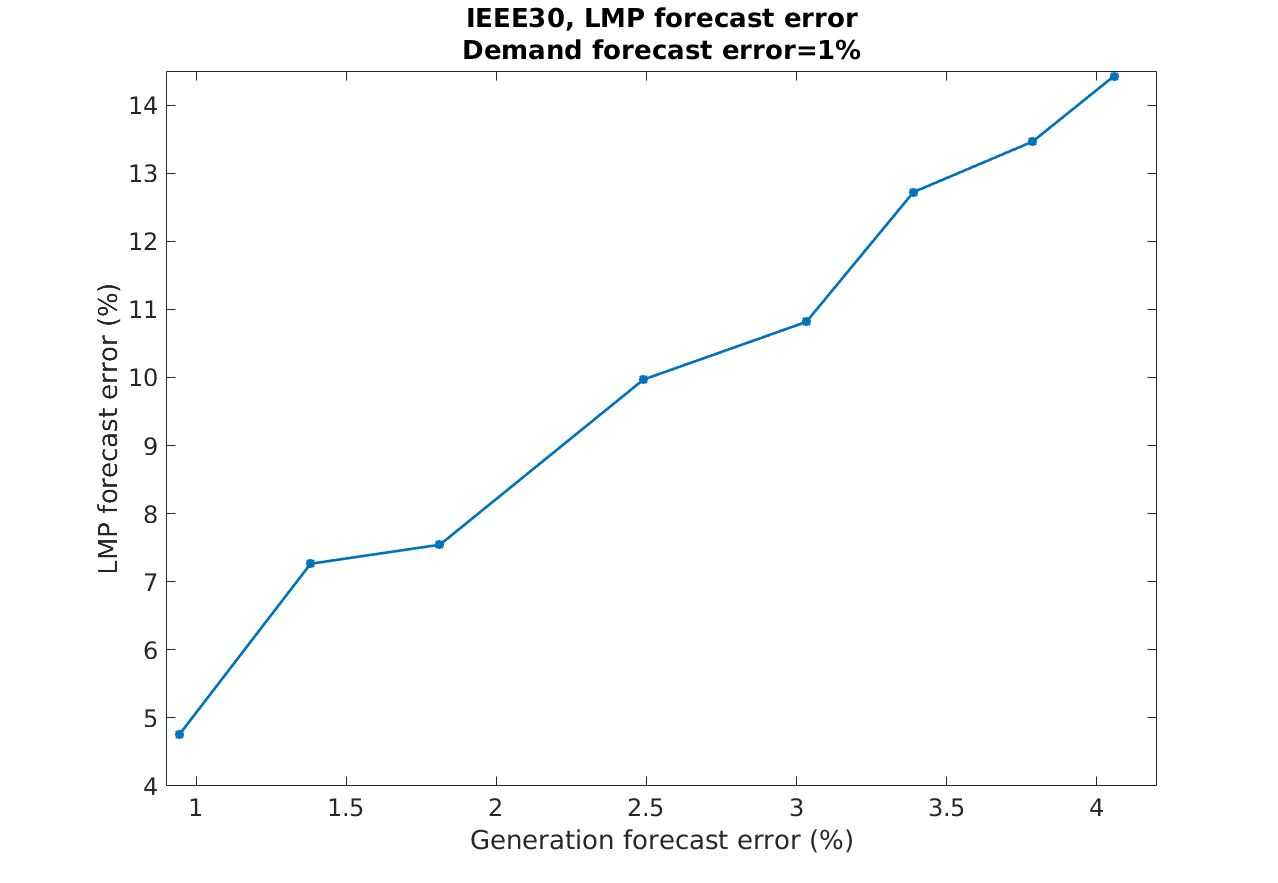}
    \caption{Demand forecast error fixed at $1\%$.}
\end{subfigure}  
\caption{LMP prediction error vs. demand (a) and renewable generation (b) forecasting error.}
\label{fig:IEEE30_error}
\end{figure}

Fig.~\ref{fig:IEEE30_error} captures the sensitivity of the predictive performance as a function of the forecasting errors in total load and renewable generation, by treating them separately. More specifically, Fig.~\ref{fig:IEEE30_error} (a) (Fig.~\ref{fig:IEEE30_error} (b)) shows how the prediction error changes as a function of the forecasting error in the total load (total renewable generation) where the generation (load) forecasting error is kept fixed. 

One of the assumptions we make is that the ratio between the nodal demand (renewable supply) and the total demand (total renewable supply) stays the same. We conduct the sensitivity analysis to evaluate how the prediction error changes as the relative errors $\frac{\|\Delta\alpha\|_2}{\|\alpha\|_2}$ and $\frac{\|\Delta\beta\|_2}{\|\beta\|_2}$ increase. The results are captured in Table~\ref{table:sensitivity}.

\begin{table}[h]
\caption{Sensitivity Analysis on nodal ratios.}
\label{table:sensitivity}
\centering
\begin{tabular}{|p{4cm}|c|c|c|c|}
\hline
Demand and generation ratio error  $\frac{\|\Delta\alpha\|_2}{\|\alpha\|_2} = \frac{\|\Delta\beta\|_2}{\|\beta\|_2}$ & 1\% & 2\% & 3\% & 4\% \\ \hline
LMP prediction error    & 6.5\% & 7.5\% & 8.5\% & 10\%  \\ \hline
\end{tabular}
\end{table}

\end{document}